\magnification 1200
\centerline{\bf Nucleosynthesis  in a Simmering Universe}
\vskip 3cm
\centerline {Daksh Lohiya, Annu Batra}
\centerline{Shobhit Mahajan, Amitabha Mukherjee}
\centerline {Department of Physics \& Astrophysics, University of Delhi,}
\centerline {Delhi 110 007, India}
\centerline {email: dlohiya@ducos.ernet.in}
\vskip 3cm
\centerline {\bf Abstract}
\vskip 1cm
     Primordial nucleosynthesis is a success story of the 
standard big bang (SBB) cosmology. We explore
nucleosynthesis  in possible models very different from SBB in which 
the cosmological
scale factor increases linearly with time right through the period during 
which nucleosynthesis occurs till the present. 
It turns out that weak interactions remain in
thermal equilibrium upto temperatures which are two orders of 
magnitude lower than the corresponding (weak interaction decoupling)  
temperatures in SBB. Inverse
beta decay of the proton ensures adequate production of helium
while producing primordial metallicity much higher than that 
produced in SBB. Attractive features of such models are the 
absence of the horizon, flatness and age problems and consistency
with classical cosmological tests. 

\vfil\eject

	Early universe nucleosynthesis is a major
``success story'' of the standard big bang (SBB) model. The 
results look rather good and the 
observed light element abundances  severely
constrain cosmological and particle physics parameters.

	Surprisingly,  a class of models 
radically different  from the standard one has a promise of
producing the correct amount 
of helium as well as the  metallicity observed in low metallicity 
astrophysical objects.
This class is defined within the Friedmann - Robertson - Walker framework
by a linear variation of the cosmological scale
factor with time.  How such an evolution can be dynamically realised shall
concern us later in this article. For the time being we outline the essential 
history 
of a  universe that is born and evolves as a Milne universe defined by
the metric:
$$
ds^2 = dt^2 - R^2(t)[{dr^2\over {(1 + r^2)}} + r^2d\theta^2 
+r^2sin^2\theta d\phi^2]
$$
with $R(t) = t$. 

	We start by summarizing the early universe nucleosynthesis
story in SBB.

	A crucial assumption in the standard model is the existence of 
thermal equilibrium at temperatures around $10^{12}K$ or $100 MeV$.
At these temperatures, the universe consists of leptons, photons 
and a contamination of baryons [mainly neutrons and protons] in thermal 
equilibrium. The ratio of weak 
reaction rates of leptons to the rate of expansion of the universe 
(the Hubble parameter) below $10^{11}K$ (age $\approx .01 $ secs) 
goes as:
$$ 
r_w \equiv {\sigma n_l\over H} \approx ({T\over 10^{10}K})^3 \eqno{(1)}
$$
The notations are as described by Weinberg [3]: 
$\sigma$ is the weak interaction
cross section, $n_l$ the density of charged leptons and $H$
the hubble parameter. 
At these temperatures, the small baryonic   
contamination begins to shift towards more protons and fewer
neutrons because of the neutron - proton mass difference. 
By $10^{10}K$ i.e. $T_9 \equiv 10$, $r_w$ falls below unity, consequently 
lepton weak interactions fall out of equilibrium and 
neutrinos  decouple. The energy distribution function 
of the neutrinos, however,
maintains a Planckian profile as the universe expands. At $ T \approx
5\times 10^{9}K$ (age of about  4 seconds), $e^+,e^-$ pairs
annihilate. The neutrinos having decoupled, all the entropy of
the $e^+, e^-$ before annihilation, goes to heat up the photons -
giving the photons a temperature which is 40\% higher than the temperature
corresponding to the neutrino Planckian profile. 
Meanwhile there is a rapid fall in the neutron production by 
electron and anti - neutrino capture by the proton and this
freezes the n/p ratio to a value slightly less than 1/5. 
This ratio now falls slowly on account of decay of
free neutrons. Further,  nuclear reactions and photo -
disintegration of light nuclei ensure a dynamic buffer of light elements
with abundances roughly determined by nuclear statistical 
equilibrium (NSE). Depending on the baryon-entropy ratio, at a critical 
temperature around $T_9 = 1$, when the n/p ratio has fallen to almost 1/7,
the deuterium concentration becomes large enough 
for efficient evolution of a whole network of reactions leading up to the 
formation of the most stable light nucleus, viz. $^4He$. This
is the characteristic temperature at which $D$ conversion
into other nuclei becomes 
a more efficient channel for the destruction of neutrons than 
neutron decay. At slightly lower 
temperatures, the deuterium depletion rate becomes small compared 
to the expansion rate of the universe [4] - resulting in residual 
abundances of deuterium and $^3He$.  Elaborate numerical codes have 
been developed [5] to describe  evolution
of the entire history from $T = 10^{11}K$ to $10^7K$. 
Abundances of deuterium, helium - 3,
helium - 4 and lithium - 7 are used to constrain the baryon -
entropy ratio, the number of light particle species in the model 
and the neutrino chemical potential.

	Taking a cue from the above narrative, we consider 
nucleosynthesis in a model in which, right through the epoch when 
$T \approx 10^{12}K$ and thereafter, the scale factor $R(t)$ increases 
as $t$ (the age of the universe). 
With such scaling, the present value of the scale parameter, i.e.
the present epoch  $t_o$, is exactly determined by the present
Hubble constant $H_o = 1/t_o$. The scale factor and the temperature of
radiation are related by  $RT \approx$ constant with effect from temperatures
$\approx 10^9K$. This follows from stress energy conservation
and the fact that the baryon - entropy ratio does not significantly
change after $kT \approx m_e$ (rest mass of the electron). 
From the present age and effective cosmic microwave background
temperature ($2.7K$), one finds the age of 
universe when $T \approx 10^{10}K$ to be around 4.5 years. Such a
universe takes some $10^3$ years 
to cool to $10^7K$. The 
rate of expansion of the universe is about $10^7$ times 
slower than corresponding rate (for the same temperature) 
in standard 
cosmology around $10^9K$. This makes a crucial [big] difference. The
first reaction would be that 
with neutron half life of 888 seconds, there would 
hardly be any neutrons left at such temperatures. Further, as we shall see
later, proton - neutron inter converting weak interaction rates remain in
equilibrium  till temperatures even below $T_9 \approx 1$. This would imply
that one would have the n/p ratio falling by the Boltzmann factor 
[$exp(-\Delta m_{n-p}/kT)$], 
again leading to a depletion of the neutron number.
``With such a low neutron count - no nucleosynthesis !!'' Indeed the standard
argument would put the issue at rest here as done recently in [18]. However,
and this is the object
of this article, such a hasty conclusion, though ``obvious'', is incorrect.
We shall see that in the conditions as stated above, the mechanism leading
to nucleosynthesis is a bit more subtle. The fact of inverse 
beta decay of the proton not freezing out, saves the day.

	To see what would happen in some detail, we start by
considering the effect of
the slow expansion on the leptonic weak interactions.	
The process of the neutrinos falling out of thermal equilibrium, for 
example, is determined by the rate of $\nu$ production per charged lepton:
$$ 
\sigma_{wk} n_l/c^6 \approx g_{wk}\hbar^{-7}(kT)^5/c^6\eqno{(2)}
$$
and the expansion rate of the universe [$H = 1/t$]. Here
$g_{wk} \approx 1.4\times 10^{-45}$ erg- cm$^3$. (We again follow
the notation and description as given by Weinberg [3])
For 
$kT > m_\mu,~ T > 10^{12}K$
$$ 
\sigma_{wk} n_l/H \approx [{T\over {1.6\times 10^{8}K}}]^4 \eqno{(3)}
$$
Here we have normalised the value of $RT \approx tT \approx$ constant
from the value $H_o \approx$ 65 km/sec/Mpc for the Hubble constant -
corresponding to $t_o \approx 15\times 10^9$ years. 
[$tT_9 \approx 2.5\times 10^9$]. Increasing 
$H_o$ even by a factor of 2 would merely lead to a change in the 
denominator on the right side of eqn.(3) to  $1.8\times 10^{8}K$.
When $kT < m_\mu$, the number density of muons is reduced by 
a factor $[exp(-m_\mu/kT)]$. Consequently, the rates of 
weak interactions involving muons get suppressed to
$$ 
\sigma_{wk} n_l/H \approx [{T\over {1.6\times 10^{8}K}}]^4 
exp[-{10^{12}K\over T}] \eqno{(4)}
$$
The corresponding rates in the standard big bang model are:
$$ 
\sigma_{wk} n_l/H \approx [{T\over {10^{10}K}}]^3 \eqno{(5)}
$$
for $kT > m_\mu$, and
$$ 
\sigma_{wk} n_l/H \approx [{T\over {10^{10}K}}]^3 
exp[-{10^{12}K\over T}] \eqno{(6)}
$$
for $kT < m_\mu$. We conclude that leptonic weak 
interactions involving muons would freeze out at temperatures
around $10^{11}K$ as in the standard model. 
However, for all leptonic 
weak interactions mediated by neutral currents
and, for weak interactions mediated by charged currents not 
involving the muons, the suppression
factor $exp[-10^{12}K/ T]$ is absent [3]. It follows that for
the following weak interactions mediated by neutral currents
$$
e^- + e^+ \longleftrightarrow {\bar \nu} + \nu
~~~~ e^\pm + \nu \longrightarrow  e^\pm + \nu
~~~~e^\pm + {\bar \nu}  \longrightarrow  e^\pm + {\bar \nu}
$$
the ratio of the reaction rates to the expansion rate $H$, for
temperatures $kT < m_e$,
would be given by:
$$ 
\sigma_{wk} n_l/H \approx [{T\over {1.6\times 10^{8}K}}]^4 
exp[-{m_e\over {kT}}] 
$$
This would
maintain the $\nu$'s in thermal equilibrium at all temperatures down 
to slightly less than $10^{9}K$. The entropy
released from the $e^+ e^-$ annihilation, at $T_9 \approx 5$, would heat up 
all the particles in equilibrium. Both neutrinos and
photons would therefore get heated up to the same temperature.
The temperature then scales by $RT =$ constant as  
universe expands. Relic neutrinos and photons 
(CMBR) would therefore have the same Planckian profile
($T \approx 2.7K$) at present. (The photon number does not
significantly change at recombination for a low enough baryon
- entropy ratio). This is in contrast
to the standard result wherein the relic neutrino temperature
is predicted to be 40\% lower than the photon temperature. 

	This has the following effect on hadronic weak decays. With the
neutrino and photon temperatures equal, the
neutron - proton weak reaction rates are given by the 
expressions [3]:
$$ 
\lambda(n\longrightarrow p) = A\int(1 - {m_e^2\over 
(Q+q)^2})^{1/2}(Q+q)^2q^2dq
$$
$$
\times (1 + e^{q/kT})^{-1}(1 + e^{-(Q+q)/kT})^{-1}\eqno{(7a)}
$$
$$ 
\lambda(p\longrightarrow n) = A\int(1 - {m_e^2\over 
(Q+q)^2})^{1/2}(Q+q)^2q^2dq
$$ 
$$
\times (1 + e^{-q/kT})^{-1}(1 + e^{(Q+q)/kT})^{-1}\eqno{(7b)}
$$
These rates have their ratio determined by the neutron - proton 
mass difference $\equiv Q \approx 15$ (with temperature measured
in units of $10^9K$):
$$
{\lambda(p\longrightarrow n)\over 
{\lambda(n\longrightarrow p}} = exp(-{Q\over T_9})\eqno{(8)}
$$
The rate of expansion of the universe at a given 
temperature being much smaller than that in the standard scenario, 
the nucleons would be in thermal equilibrium till temperatures
slightly below $10^9K$.  $X_n \equiv$ 
the ratio of neutron number to the neutron plus proton number is
given by:
$$ 
X_n = {\lambda(p\longrightarrow n)\over 
{\lambda(p\longrightarrow n) + \lambda(n\longrightarrow p)}}
= [1 + e^{Q/T_9}]^{-1}\eqno{(9)}
$$

	Let us now re-state the problem at hand. 
A universe which evolves according to $R(t) = t$ is some
tens of years old at temperatures $T_9 \approx 1$ and the neutron - 
proton ratio at such temperatures keeps falling as 
$n/p \approx exp(-15/T_9)$.
One is tempted to naively ask: were nucleosynthesis  to
commence at temperatures below $T_9 =1$, then as (1) the age of universe 
is much larger than the neutron life time and (2) with 
the $n/p$ ratio reaching very 
low levels, why would there be any significant nucleosynthesis ? 

	First of all,
as long as n's and p's are held in equilibrium by weak interactions,
the age of the universe being large as compared to the neutron lifetime 
has no effect on the $n/p$ ratio. As long as neutron - proton inter
conversion rates are large as compared to the rate of expansion of 
the universe, $X_n$ is given by eqn(9). Secondly, 
the low level of $n/p$ at the time when nucleosynthesis commences does
not on its own determine the amount of heavier elements produced. If existing
neutrons at any given stage are removed to branch off to the nucleosynthesis
network, then, as weak interactions are still in equilibrium, inverse 
beta decay of the protons by electron capture would restore and maintain
the $n/p$ ratio to its equilibrium value. This is similar to an analogous
situation in chemical kinetics referred to as the ``law of mass action''. 
Given an equilibrium buffer of reactants and products, if any of the
reactants or the products are removed, the reaction proceeds to restore the 
equilibrium concentrations. This operates
particularly if the precipitation of the reactants
or the products is at a rate that is smaller than the relaxation time 
of the equilibrium reaction [19]. We shall come to this point after 
demonstrating the crucial role that inverse beta decay can play in this
slow evolution.
$Assuming$, for instance, that nucleosynthesis were to commence at
a sharply defined temperature around $T_9 \approx 1$, 
one sees from eqn(9) that   
there is hardly any concentration of neutrons at this temperature.
However weak interactions
have not frozen off and inverse beta decay can still convert protons 
into neutrons. If the ratio of number of protons 
that convert into neutrons after this epoch, to the total baryon number of
the universe is roughly 1/8, and all the neutrons so created 
were to branch into 
the nucleosynthesis channel with $100\%$ efficiency, we could get the observed
$\approx 25\% ~^4He$.
We can constrain the temperature $T_{9o}$ at which
nucleosynthesis ought to commence in such a case. Eqn(8) implies:
$$
{{\lambda(p\longrightarrow n)}\over {\lambda(n\longrightarrow p)}}
= exp(-{Q\over kT})\approx e^{-15/T_9}\eqno{(10)}
$$
If $\tau$ is the neutron life time viz. $\lambda(n\longrightarrow p)^{-1}$ 
at low
temperatures, eqn(10) gives the following equation for the proton ratio:
$$
{\dot X_p} \approx - {1\over \tau}e^{-15/T_9}X_p \eqno{(11)}
$$
This is exactly integrated, starting from a temperature $T_{9o}$, 
to give:
$$
X_p \approx X_{po}exp[-{10^9\over {15\tau}}e^{-15/T_{9o}}]
$$
$X_{po} - X_p$ is the number of protons converted to neutrons. 
If all the neutrons thus produced were to precipitate as
$^4He$ as described above, the amount of helium is just:
$$
Y_{He} \approx 2[1 - exp[-{10^9\over 15\tau}e^{-15/T_{9o}}]]\eqno{(12)}
$$
This is  $\approx 24\%$  for $T_{9o} \approx 0.9$.

	In the above analysis, we assumed that nucleosynthesis
commenced at a sharply defined temperature and all neutrons formed thereafter
branched off into the element production channel. A $100\%$ branching into 
nucleosynthesis would never be achieved and one has to run a full
numerical code as described later.
The example suffices to point out that the naive analysis [18] is not correct
and one has to proceed 
with caution. Weak interactions being in equilibrium and slow rate of 
expansion of the universe, contribute to salvage nucleosynthesis. 

	Basically, with a judicious choice of the baryon entropy ratio, 
one can remove neutrons from the equilibrium buffer consisting of neutrons,
protons, deuterium and photons at a rate smaller than the relaxation period
of the buffer. Inverse beta decay would keep replenishing neutrons into the
buffer. The rate of heavier element production would be slow - but  steady.
One has hundreds of years at one's 
disposal to have the total helium add up to
the right required amount.

	We proceed to outline a clearer
picture of what actually happens in such a slow evolution.
The baryonic content of the universe at temperatures below
$T_9 \approx 10$ consists of protons and  neutrons and a buffer of 
light elements primarily consisting of deuterium [6].
$$
X_{^2D} \approx X_nX_pexp[25.82/T_9]10^{-5}T_9^{1.5}\eta
$$ 
Here $\eta$ is the baryon entropy ratio.
As long as the 
rate of deuterium depletion into heavier elements is much smaller than
that of $n[p,D]\gamma$ and the reverse reactions, $^2D$ would 
be maintained near the above
equilibrium value obtained by the detailed balancing of
the $n[p,D]\gamma$ reaction. Nucleosynthesis can proceed
by the following reversible reactions:
$$
n[p,D]\gamma;~n[D,^3H]\gamma;~n[^3He,^4He]\gamma;~n[^3He,^3H]p;~
p[D,^3He]\gamma;~p[^3H,^4He]\gamma;~
$$
$$
D[D,^3He]n;~D[D,^3H]p;~D[^3H,^4He]n;~D[^3He,^4He]p;~^3He[^3He,^4He]2p
$$
Rates of all these reactions are listed in several review articles eg. [5,6]. 
At the temperatures of interest, 
the reverse reactions for all but $n[p,D]\gamma$ are severely suppressed
in comparison to the forward reaction rates. Thus 
small amounts of $^4He$ would keep 
precipitating out of the network. The most important point that enables 
sufficient nucleosynthesis to occur is that the 
suppression of the forward reaction
rate on account of low $D$ abundance is compensated by the large amount of
time for which the universe holds at these temperatures. This is in marked 
contrast to the situation in 
SBB where the universe holds at these temperatures
for just a few seconds. As the equilibrium value of
$D$ is sensitive to the baryon entropy ratio, so would the element production
be. Getting the right amount of $^4He$ translates 
into an appropriate requirement on the baryon-entropy ratio. 
In SBB one may recall that at temperatures even higher than
the so called $D$ photodissociation bottleneck, any heavier elements
formed would survive. However, and this is the most important point, 
in SBB the universe holds at these higher temperatures for a very short time
and hardly any production has taken place by the time the above bottleneck
is arrived at. In the case at hand, nucleosynthesis starts around 
$T_9 \approx 7$ at a very small rate. However, the long period for which the 
universe  holds more than compensates for the small rate of production.
The rate of production of 
helium is smaller than the inverse beta decay rate of
the proton !! This ensures that all neutrons branching off to the 
nucleosynthesis channel are compensated by more being formed by the 
inverse beta decay.

	Fortunately one has an extremely user friendly code [5]
that we modified 
to suit the taxing requirements of the much stiffer rate equations
that we encounter in our slowly evolving universe. To get 
convergence of the rate equations  for 26 nuclides and a network of
88 reactions (as given in Kawano's code), we were forced to
rewrite essential subroutines in quadruple precision. 
The code incorporates the variation of the baryon entropy ratio
during the electron positron annihilation epoch.
The results for different values of final baryon entropy ratio 
$\eta$ are shown in
table I.  We find consistency with the $^4He$ abundances for
$\eta \approx 10^{-8} $. The metallicity produced is 8 orders of 
magnitude greater than the corresponding value one gets in the
early universe in the Standard model. This is
also a consequence of the slow expansion in this model that allows for
more time for reactions that build up metallicity. A 
locally higher $\eta$ in an inhomogeneous model can further enhance
metallicity. 

	We would like to add here that when we first addressed ourselves
to this problem of obtaining the right amount of helium in an $R(t) = t$
cosmology, we had realized that the inverse beta decay would play
a vital role. One has just one parameter, the baryon entropy
ratio, to be varied in a hope that one keeps precipitating neutrons
in the form of helium at a rate small in comparison to the relaxation time 
of equilibrium of the buffer. Yet the rate ought not to be so small 
that even over the long period that the universe takes to cool, significant
nucleosynthesis is not achieved. 
We varied $\eta$ in our numerical code in search of a value that
would yield the right amount of helium.
We would have regarded the model as 
non - viable had our search yielded a very large
(say $10^{-6}$) or an extremely small (say $10^{-11}$) value for $eta$.
Our search yielded a value $10^{-8}$ which
is the kind of value that is sought for eg. in Weinberg's classic [3]. We 
find this quite encouraging.

    To get the observed abundances of light elements besides $^4He$, 
one would have to fall back upon a host of other mechanisms 
that were being explored in the SBB in the pre - 1976 days.
The most popular processes are: (i) nucleosynthesis by secondary 
explosions of super massive objects [6], (ii) nucleosynthesis in 
inhomogeneous models, (iii) effect of inhomogeneous $n/p$ ratios
as the universe comes out of the QGP phase transition, (iv)
spallation of light nuclei at a much later epoch.
It is easy to rule out the survival of $D$ by the processes (ii) 
and (iii) while the process (i) requires very special initial 
conditions. It also shares a common difficulty with process (iv),
viz.: the production of $D$ to the required levels is
possible but it is accompanied by an overproduction of lithium.
Any later destruction of lithium in turn completely destroys
$D$. Within the framework of the cosmological evolution that
we are exploring here, we find the best promise in a model that
would combine (ii) and (iv). Table 1 displays the extreme
sensitivity of $^4He$ production to $\eta$. In an inhomogeneous model
with a spatially varying $\eta$, there would hardly be any 
$^4He$ production in a region with $\eta$ lower by (say) a  
factor of two. Thus we can have proton rich clouds 
in low density regions and $^4He$ and metal rich clouds in
the higher density regions produced as the universe cools 
from $T_9 \approx 5$. The spallation of the former
on the latter, at a subsequent (cooler) epoch, would produce 
$D$ without the excess production of lithium [7].

	We feel that one should be able to dynamically account
for such conditions within the framework of models we 
outline in the conclusion.
\vskip 1cm
\centerline{\bf Conclusion}

      The purpose of the article is to show that the class of FRW
cosmological models where the scale factor grows linearly with time 
cannot be trivially discarded away on account of
SBB nucleosynthesis constraints.
In any model in which the rate of expansion of the universe is low
enough, inverse beta decay remains in equilibrium and
can lead to adequate $^4He$ and metal production. Further, in
principle, it is possible to produce $D$ by spallation of hydrogen
rich clouds over a $^4He$ - metal rich medium at a later epoch.

       One may well ask: (1) Does $R(t) = t$ coasting lead to a viable 
cosmology  ? 
and, a related question: (2) How could such an evolution be theoretically
realised in a gravity model ?

	As regards the first query: 
An FRW metric with $R(t) = t$ has interesting features. It has no horizon
problem. At any given time $t > 0$ every observer can see the entire 
universe. Further, as shown below, there are models in which such a coasting is
independent of any ``critical density''. Thus the metric does not suffer the
flatness problem. Classical cosmological tests, namely:
the Hubble diagram (luminosity distance-redshift 
relation), the angular diameter distance - redshift relation and the galaxy 
number count-redshift relations do not rule out such  
a ``coasting'' cosmology [8,9,18]. In fact the best fit for the
latest observations on type IA supernovae [16] is practically indistinguishable
from that expected of a $R(t) = t$ [$\Omega_M = \Omega_\Lambda = 0$]
cosmology. 
The age of universe inferred from a measurement of the Hubble 
parameter is $t_o = 1/H_o$ and is comfortably concordant with the age 
estimates of the oldest objects [clusters, low metallicity clouds etc.].
Finally, the low metallicity that one sees in type II objects poses
a problem in SBB. There is
no object in the universe that has quite the abundance [metallicity]
of heavier elements as is produced in the 
``first three minutes'' (or so)
in SBB. One relies on some kind of
re - processing, much later in the history of SBB, 
to get the low observed  
metallicity in [eg.] old clusters and inter - stellar clouds. This could
[for instance] be in the form of a generation 
of very short - lived type III stars. Large
scale production and recycling of metals through such exploding early
generation stars leads to verifiable observational 
constraints. Such stars would be visible as 27 - 29 magnitude stars
appearing any time in every square arc - minute of the sky.
Serious doubts have been expressed on the existence and
detection of such signals [1]. In the nucleosynthesis model described
in this article, the primordial metallicity obtained is quite close to  
lowest reported metallicity. An $R(t) = t$ cosmology comes with 
characteristic predictions. A vanishing deceleration parameter and
the equality of effective temperatures of relic microwave background
and neutrinoes being two of them.

	 Of late [2],   
observations have suggested the need for a careful scrutiny and a 
possible revision of the status of SBB nucleosynthesis from reported
high abundance of $D$ in several $Ly_\alpha$ systems. Though the 
status of these observations is still a matter of debate, and  
(assuming their confirmation), attempts to
reconcile the cosmological abundance of deuterium and the number of 
neutrino generations within the framework of SBB are still on, we feel that
alternative scenarios should be explored.

	 We finally address the second query i.e. the issue 
of realising linear evolution 
within the framework of a Friedman cosmology. Within conventional
Einstein's theory, such an evolution
can be accounted for in a universe dominated by a hypothetical
`K - matter' [8] for which 
the density scales as $R^{-2}$. 
However, if one requires
this matter to dominate even during the nucleosynthesis era, the K -
matter would almost close the universe. There would hardly be any
baryons in the present epoch. One has to look elsewhere. 
Fortunately, linear evolution of the scale factor is a 
generic feature of a large class of non - minimally coupled theories [12, 17].
These are models in which a non - minimal 
coupling diverging with time is used
to dynamically scale the cosmological constant to zero. Such an  
evolution of the scale factor is also possible in alternative effective
gravity and higher order gravity theories. Ellis and Xu [10] for example,
consider a higher order gravity theory with action:
$$ 
S = \int d^4x\sqrt{-g}[\alpha R^2 - \beta R] \eqno{(14)}
$$
in the weak field approximation, the effective Newtonian potential
is:
$$\phi = - {a\over r} + b{exp(-\mu r)\over r} \eqno{(15)}$$
For $\mu r << 1$ we can have a canonical effective attractive theory. 
Over 
large distances, the effective potential is 
dominated by the first repulsive term
alone. A similar possibility occurs
in the conformally invariant higher order theory of gravity[11].
Choosing the gravitational action to be the square of the
Weyl tensor gives rise to an effective gravity action:
$$ 
S = \int d^4x\sqrt{-g}[\alpha C^2 - \beta R]\eqno{(16)}
$$
The dynamics of a conformally flat FRW 
metric is driven by the anomalous repulsive term  $\beta R$ alone. The 
FRW - scale factor in such a cosmology approaches linear evolution
at large cosmic time. Canonical
attractive domains occur in the model as non - conformally flat 
perturbations in the FRW spacetime.

    Linear evolution of the scale factor would also be possible
in the following ``toy'' model [13] that combines the Lee - Wick
construction of non - topological soliton [NTS] solutions [14] in
a variant of an effective gravity model proposed by Zee [15]. 
Consider the action:
$$
S = \int d^4x\sqrt{-g}[U(\phi)R + 
{1\over 2}\partial_\mu\phi\partial^\mu\phi - V(\phi) + L_m] \eqno{(17)} 
$$
Here $\phi$ is a scalar field non - minimally coupled to the
scalar curvature through the function $U(\phi)$, $V(\phi)$ its 
effective potential and $L_m$ the matter field action. $L_m$
includes a Higgs coupling of $\phi$ to a fermion. Let $V$ have a 
minimum at $\phi_{min}$ and a zero at $\phi^o$. We also choose
the Higg's coupling such that the effective fermion mass at
$\phi = \phi_{min}$ is greater than the effective fermion mass
at $\phi = \phi^o$. Finally we choose the non - minimal 
function $U(\phi_{min}) >> U(\phi^o)$. These conditions are
sufficient for the existence of large NTS's with the scalar
field trapped at $\phi = \phi^o$ in the interior of a large 
ball and quickly going to $\phi = \phi_{min}$ across the surface
of the ball. With a judicious choice of the surface tension, 
these balls could be larger than  typical halos
of galaxies. The interior and exterior of such a ball 
would be regions with
effective gravitational constant 
$[U(\phi^o)]^{-1}~\&~[U(\phi_{min}]^{-1}$ 
respectively. With $[U(\phi_{min})]$ large enough, the universe 
would evolve as a curvature dominated universe [without
any `K - matter'].

\vskip 2cm

Acknowledgment: Helpful discussions with Jim Peebles, 
T. W. Kibble and G. W. Gibbons
are gratefully acknowledged.
\vskip 1cm 
	Copy of the numerical code is available from the authors. The
executable file would need an architecture that would support quadruple 
precision calculation. 
\vfil\eject

\centerline{\bf References}

\item{1.} J. M. Escude \& M. J. Rees, Astro-ph 9701093 (1997) 
\item{2.} G. Steigman,  Astro-ph/9601126 (1996)
\item{3.} S. Weinberg, ``Gravitation and Cosmology'', John Wiley \&
sons, (1972).
\item{4.} See for example E. W. Kolb, and M. S. Turner, 
``The Early Universe'', Addison Wesley (1990).
\item{5.} L. Kawano, (1988), FERMILAB - PUB -88/34 and references therein.
\item{6.} R. V. Wagoner, Ap. J. Supp., $\underline{162}$, 18, 247 (1967)
\item{7.} R. I. Epstein, J. M. Lattimer \& D. N. Schramm, Nature
$\underline{263}$, 198 (1976); R. Epstein, Astrophys. J. 
$\underline{212}$, 595 (1977); F. Hoyle \& W. Fowler, Nature
$\underline{241}$, 384 (1973)
\item{8.} E. W. Kolb, Astrophys. J., ${\underline 344}$, 543 (1989)
\item{9.} M. Sethi \& D. Lohiya,  ``Aspects of a coasting universe'',
GR15 proceedings (1997)
\item{10.} G. F. R. Ellis and M. Xu, 1995 (private communication).
\item{11.} P. Manheim and D. Kazanas, Gen. Rel \& Grav.
$\underline {22}$ 289 (1990)
\item{12.} H. Dehnen \& O. Obregon, Ast. and Sp. Sci. $\underline{17}$,
338 (1972); ibid. $\underline{14}$, 454 (1971)
\item{13.} D. Lohiya \& M. Sethi, ``A program for a problem free
cosmology within a framework of a rich class of scalar tensor
theories'' Class. and Quan. Gravity (1999), to be published
\item{14.}(a) T. D. Lee, Phys. Rev. $\underline{D35}$, 3637 (1987);
T. D. Lee \& Y. Pang, phys. Rev. $\underline{D36}$, 3678 (1987);
(b) B. Holdom, Phys. Rev. $\underline{D36}$, 1000 (1987)
\item{15.} A. Zee, in ``Unity of forces in nature''
Vol II, ed. A. Zee, P 1082, World Scientific (1982);
Phys. Rev. Lett., $\underline{42}$, 417 (1979); 
Phys. Rev. Lett., $\underline{44}$, 703 (1980).
\item{16.} S. Perlmutter et al, ``Measurement of $\Omega$ and $\Lambda$
from 42 high redshift objects; Astro-ph/9812133
\item{17.} A. D. Dolgov in ``The very early universe'', ed. G. W. Gibbons,
S. W. Hawking, S. T. Siklos, Cambridge university press, 1982.
L. H. Ford, Phys. Rev. $\underline{D35}$, 2339, 1987
\item{18.} M. Kaplinghat, G. Steigman, I. Tkachev and T. P. Walker,
astro-ph/9805114
\item{19.} J. H. Noggle, ``Physical Chemistry'', Pubs. Little Brown and Co.,
Boston 1985. 
\vfill
\eject

\vskip 3cm

\centerline {\bf {TABLE I}}
\vskip 1cm
\centerline{Abundances of Some Light Elements and Metals.} 

\vskip 2cm

\settabs 7 \columns
\+\bf$\eta$&$\bf^2 H$&$\bf^3 H$&
$\bf^3 He$&$\bf^4 He$&$\bf^7 Be$&$\bf^8 Li$ \& above \cr
\smallskip
\+$(10^{-9})$&$(10^{-18})$
&$(10^{-25})$&$(10^{-14})$&$(10^ {-1})$&$(10^{-11})$&$(10^{-8})$ \cr
\smallskip
\+$9.0$&$2.007$&$1.25$&$8.65$&$2.03$&$1.39$&$8.06$ \cr
\+$9.1$&$2.008$&$1.26$&$8.63$&$2.06$&$1.32$&$8.63$ \cr
\+$9.2$&$2.009$&$1.26$&$8.60$&$2.10$&$1.23$&$9.35$ \cr
\+$9.3$&$2.010$&$1.27$&$8.59$&$2.11$&$1.19$&$9.75$ \cr
\+$9.4$&$2.014$&$1.26$&$8.56$&$2.15$&$1.11$&$10.66$ \cr
\+$9.5$&$2.015$&$1.27$&$8.50$&$2.18$&$1.05$&$11.41$ \cr
\+$9.6$&$2.016$&$1.28$&$8.52$&$2.19$&$1.01$&$11.88$\cr
\+$9.7$&$2.017$&$1.28$&$8.49$&$2.22$&$0.96$&$12.69$ \cr
\+$9.8$&$2.020$&$1.29$&$8.47$&$2.25$&$0.91$&$13.51$ \cr
\+$9.9$&$2.020$&$1.29$&$8.45$&$2.28$&$0.86$&$14.47$ \cr
\+$10.0$&$2.020$&$1.30$&$8.43$&$2.30$&$0.83$&$15.19$ \cr

\bigskip
\noindent
\+Initial Temperature   $10^{11}K$ \cr
\+Final Temperature  $10^7K$ \cr

\bye